\documentclass[sigplan]{acmart}
\AtBeginDocument{%
  \providecommand\BibTeX{{%
    \normalfont B\kern-0.5em{\scshape i\kern-0.25em b}\kern-0.8em\TeX}}}

\acmYear{2021}\copyrightyear{2021}
\setcopyright{acmlicensed}
\acmConference[PPoPP '21]{26th ACM SIGPLAN Symposium on Principles and Practice of Parallel Programming}{February 27--March 3, 2021}{Virtual Event, Republic of Korea}
\acmBooktitle{26th ACM SIGPLAN Symposium on Principles and Practice of Parallel Programming (PPoPP '21), February 27--March 3, 2021, Virtual Event, Republic of Korea}
\acmPrice{15.00}
\acmDOI{10.1145/3437801.3441620}
\acmISBN{978-1-4503-8294-6/21/02}

\usepackage{tabularx}
\usepackage{booktabs}
\usepackage{subcaption}
\usepackage{xspace}
\usepackage{mathtools}
\usepackage{algorithm}
\usepackage[noend]{algpseudocode}

\newcommand{\CC}{C\nolinebreak\hspace{-.05em}\raisebox{.4ex}{\tiny\bf +}\nolinebreak\hspace{-.10em}\raisebox{.4ex}{\tiny\bf +}}
\def\CC{{C\nolinebreak[4]\hspace{-.05em}\raisebox{.4ex}{\tiny\bf ++}}}

\newcommand{\gathercoll}{Gather\xspace}
\newcommand{\allgather}{Allgather\xspace}
\newcommand{\scatter}{Scatter\xspace}
\newcommand{\reduce}{Reduce\xspace}
\newcommand{\broadcast}{Broadcast\xspace}
\newcommand{\reducescatter}{Reducescatter\xspace}
\newcommand{\allreduce}{Allreduce\xspace}
\newcommand{\alltoall}{Alltoall\xspace}

\newcommand{\dgxone}{DGX-1\xspace}

\newcommand{\amd}{AMD\xspace}

\newcommand{\chunkstep}[2]{$#1$-chunk $#2$-step\xspace}

\newcommand{\tool}{SCCL}
\newcommand{\toollong}{Synthesized Collective Communication Library}

\newcommand{\setwhere}{\;|\;} 
\newcommand{\qst}{\;} 
\newcommand{\powerset}{\mathcal{P}}
\newcommand{\posint}{\mathbb{Z}_{\geq0}}

\newcommand{\range}[1]{\left[#1\right]}
\newcommand{\dotscend}{\dotsc\hspace{-0.08em}}

\newcommand{\Ite}[3]{\mathrm{ITE}(#1,#2,#3)}

\newcommand{\collectiveproblem}{\textsc{SynColl}\xspace}

\newcommand{\size}{P}

\newcommand{\pre}{\mathit{pre}}
\newcommand{\post}{\mathit{post}}
\newcommand{\chunk}{C}
\newcommand{\gchunk}{G}
\newcommand{\toglobal}{\mathit{ToGlobal}}
\newcommand{\steps}{S}
\newcommand{\rounds}{R}

\newcommand{\bw}{B}
\newcommand{\rparts}{Q}
\newcommand{\sends}{T}
\newcommand{\start}[2]{\mathit{time}_{#1,#2}}
\newcommand{\send}[3]{\mathit{snd}_{#1,#2,#3}}
\newcommand{\qouta}{\mathit{Q}}

\newcommand{\broadcasting}{non-com\-bin\-ing\xspace}
\newcommand{\broadcastingCap}{Non-com\-bin\-ing\xspace}
\newcommand{\reducing}{com\-bin\-ing\xspace}
\newcommand{\reducingCap}{Com\-bin\-ing\xspace}

\newcommand{\etal}{\textit{et al}.}

\begin{document}

\title{Synthesizing Optimal Collective Algorithms}


\author{Zixian Cai}
\authornote{Both authors contributed equally to the paper. 
The work was done during internships at Microsoft Research.} 

\affiliation{
  \department{Research School of Computer Science}       
  \institution{Australian National University}           
  \city{Canberra}
  \state{ACT}
  \country{Australia}                   
}
\email{zixian.cai@anu.edu.au}          

\author{Zhengyang Liu}
\authornotemark[1]
\affiliation{
  \department{School of Computing}              
  \institution{University of Utah}            
  \city{Salt Lake City}
  \state{UT}
  \country{USA}                    
}
\email{liuz@cs.utah.edu}          

\author{Saeed Maleki}

\affiliation{
  \institution{Microsoft Research}           
  \city{Redmond}
  \state{WA}
  \country{USA}                   
}
\email{saemal@microsoft.com}         

\author{Madanlal Musuvathi}

\affiliation{
  \institution{Microsoft Research}           
  \city{Redmond}
  \state{WA}
  \country{USA}                   
}
\email{madanm@microsoft.com}         

\author{Todd Mytkowicz}
\affiliation{
  \institution{Microsoft Research}           
  \city{Redmond}
  \state{WA}
  \country{USA}                   
}
\email{toddm@microsoft.com}         

\author{Jacob Nelson}
\affiliation{
  \institution{Microsoft Research}           
  \city{Redmond}
  \state{WA}
  \country{USA}                   
}
\email{jacob.nelson@microsoft.com}         

\author{Olli Saarikivi}

\affiliation{
  \institution{Microsoft Research}           
  \city{Redmond}
  \state{WA}
  \country{USA}                   
}

\email{olsaarik@microsoft.com}         
%

\renewcommand{\shortauthors}{Zixian Cai, Zhengyang Liu \etal}

\begin{abstract}
Collective communication algorithms are an important component of
distributed computation.  Indeed, in the case of deep-learning,
collective communication is the Amdahl's bottleneck of data-parallel
training.   

This paper introduces \tool{} (for Synthesized Collective
Communication Library), a systematic approach to synthesizing collective
communication algorithms that are explicitly tailored to a particular
hardware topology.  \tool{} synthesizes algorithms along the
Pareto-frontier spanning from latency-optimal to bandwidth-optimal
implementations of a collective.  The paper demonstrates how to encode
the synthesis problem as a quantifier-free SMT formula which can be
discharged to a theorem prover. We show how our carefully built encoding enables
\tool{} to scale.

We synthesize novel latency and bandwidth optimal
algorithms not seen in the literature on two popular hardware
topologies. We also show how \tool{} efficiently lowers algorithms to
implementations on two hardware architectures (NVIDIA and AMD) and
demonstrate competitive performance with hand optimized collective
communication libraries. 

\end{abstract}




\maketitle

\section{Introduction}
\begin{figure}
\includegraphics[page=1,width=\columnwidth]{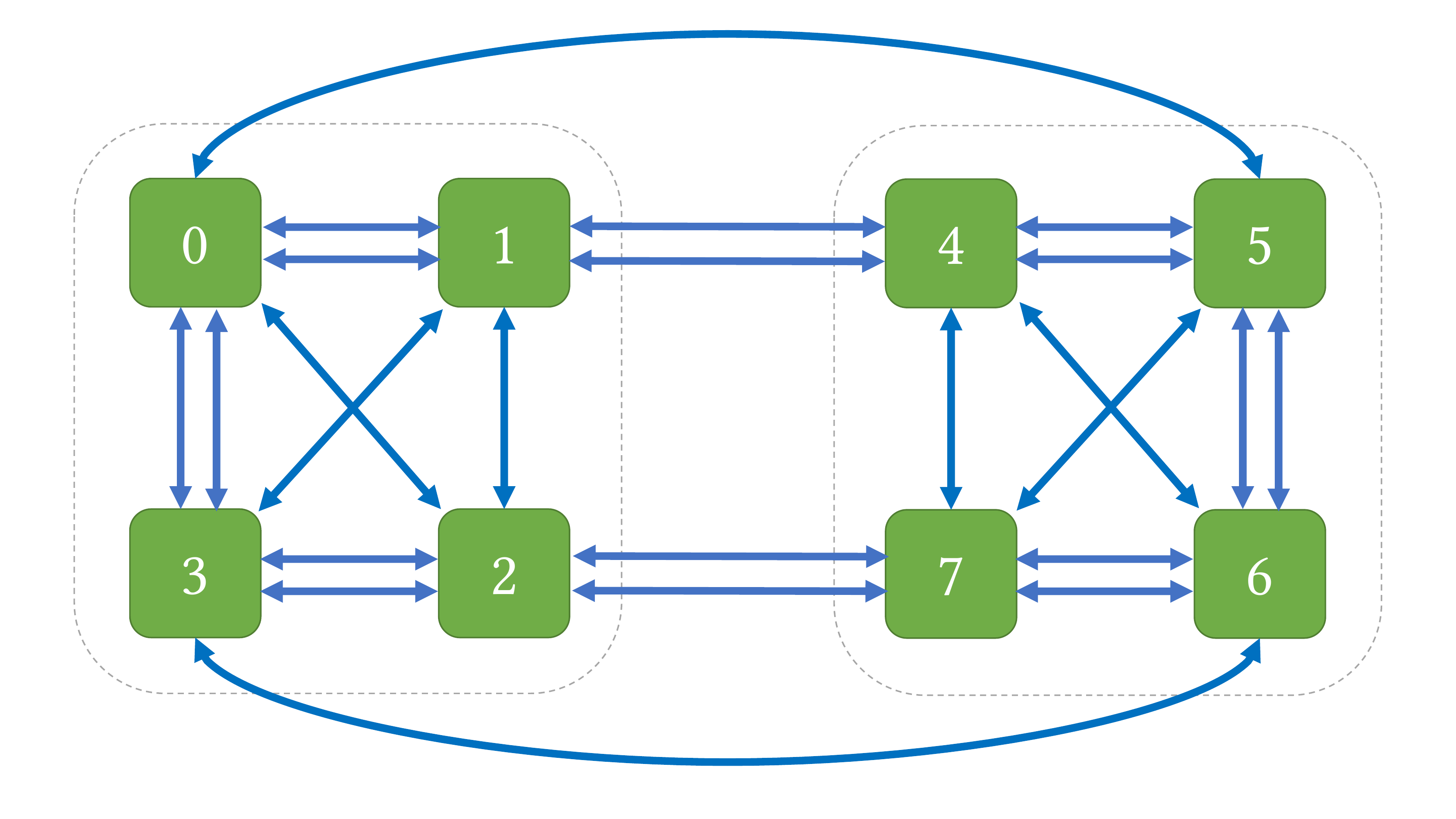}
\caption{NVLink topology of an NVIDIA DGX-1.}
\label{fig:dgx1-topo}
\end{figure}

Recent trends in machine learning towards training and serving large models together with the stagnation of Moore's-law-induced compute performance has led system designers to include novel high-bandwidth interconnect networks both within and across nodes in distributed clusters. For instance, a \dgxone server consists of two x86 processors and eight GPUs, interconnected by NVIDIA's NVLink network as shown in Figure~\ref{fig:dgx1-topo}. These networks' designs are motivated as much by the need to perform efficient \allreduce, a crucial primitive in machine learning, as well as by hardware considerations such as signal integrity, cooling and physical layout. 
A wide variety of similar accelerators with novel high-speed interconnects are used to train machine learning models today, including AMD's MI50 GPUs~\cite{mi50}, Graphcore's IPUs~\cite{graphcore} and Google's TPUs~\cite{tpu}.

These novel topologies require novel communication kernels to maximize performance. Today these kernels are written and optimized manually. For instance, NVIDIA Collective Communication Library (NCCL) has two general algorithms for the supported operations such as \allreduce: a high-bandwidth ring algorithm and a low-latency tree algorithm. These implementations are manually written and they do not necessarily have the best performance for different topologies including \dgxone's. On one hand, repeating this manual effort for other communication primitives such as \alltoall or extending already implemented algorithms to a wide variety of hardware topologies is simply infeasible. 

On the other hand, optimizing these communication kernels for performance for each topology and buffer size is crucial. For instance, we found 30\% of the training time for the 8.3 billion parameter Megatron language model with model parallelism is spent inside \allreduce where each
buffer is of medium size (10-100MB). Also, for data parallelism, the communication buffers
could range from a few KBs (one layer) to a few GBs (the entire model).
We expect this wide range of sizes as large models are developed and trained on 
larger distributed clusters.

In this paper, we automatically synthesize high-performance communication kernels. 
Given a topology, specified as a graph with bandwidth constraints on nodes and edges, and a communication primitive, specified as the pre- and post-condition on data location and computation on it, we generate~(Section~\ref{sec:synthesis}) a quantifier-free SMT formula that captures the set of all feasible algorithms that implement the primitive on the input topology.
Exploring this space to appropriately minimize the number of communication steps or decrease the granularity of communication at each step, is a computationally difficult problem. We exploit
an SMT solver to synthesize algorithms that explore this tradeoff along the Pareto frontier between latency-optimality and bandwidth-optimality.
For every solution from the SMT solver, we automatically generate and lower~(Section~\ref{sec:lowering}) high-performance implementations.


When using SMT, finding the right encoding can make all the difference for the
feasibility of an approach. This paper details the important design choices in
our encoding that help it scale to all of our hardware targets. We use the SMT
encoding for \broadcasting collectives, such as \broadcast, while for \reducing
collectives, such as \reduce, we employ a reduction back to the synthesis
problem for \broadcasting collectives.
This reduction generalizes a well known fact that some \reducing collectives may
be produced by inverting a \broadcasting one, e.g. \reduce by inverting \broadcast.





We implement our approach in a tool called \toollong{}~(\tool{}),
which probes the target hardware topology, synthesizes algorithms for
it using Z3~\cite{z3} and finally generates CUDA code that efficiently implements that algorithm.  These algorithms are
synchronous; at every step of the algorithm, one or more of the nodes
send and/or reduce data from others.

Some of the algorithms we synthesize are novel, with no known counterparts in
the literature occupying the same latency-bandwidth tradeoff. For example, we
have produced a latency-optimal 2-step (4-step) algorithm for 
the \allgather (\allreduce) primitive in the DGX-1 topology (Figure~\ref{fig:dgx1-topo}) and
a bandwidth-optimal 3-step (6-step) algorithm for the \allgather (\allreduce) primitive on the 
same topology.  In addition to providing novel
algorithms, our approach informs us when a combination of bandwidth and number
of steps is \emph{not possible}. This makes our synthesis approach a tool for
probing the algorithmic properties that a given topology provides, which is
useful for co-design of hardware interconnects with communication libraries.
Our evaluation~(Section~\ref{sec:evaluation}) shows us that this approach scales and beats NCCL in almost all cases.

To summarize, the contributions of our paper are as follows:
\begin{itemize}
    \item A formalization of the synthesis problem for \broadcasting collectives.
    \item A general strategy for encoding the synthesis problem for
    collective communications algorithms into the quantifier-free linear integer
    arithmetic (QF\_LIA) sub-logic of the SMT-LIB logic.
    \item A reduction from the synthesis problem for \reducing collectives to that for \broadcasting collectives.
    \item A description of how \tool{} generates efficient code for the algorithms we synthesize on nodes with NVIDIA or AMD GPUs.
    \item An evaluation of \tool's generated algorithms on common server topologies for deep learning workloads and a comparison against NCCL.
\end{itemize}



\section{Overview}
This section provides an overview of synthesizing latency- and bandwidth-optimal algorithms, using \allgather for the 
\dgxone topology~(Figure~\ref{fig:dgx1-topo}) as the running example. 

\subsection{Collective Communication Primitives}
\label{sec:background-collectives}
Collective communication primitives allow nodes in a networked system to perform operations on shared data. As an example, if each node has some input data, the \allgather primitive transfers these data to all of the nodes.  One way to implement this is for each node to independently send its data to all other nodes. But, an algorithm in which the nodes collectively work together can be more efficient. The efficiency of such algorithms depends on the network topology.

\subsection{Topology}
The network topology specifies how the nodes are connected with each other and the latency and bandwidth constraints on the links connecting them. Consider the \dgxone topology shown in Figure~\ref{fig:dgx1-topo}. It consists of $8$ GPUs (or nodes, in the above formalism) split into two groups $\{0,1,2,3\}$ and $\{4,5,6,7\}$. The nodes in each group are fully connected. In addition, there are four inter-group links as shown in the figure. These nodes are connected through 
NVLinks, with some nodes connected with two parallel NVLinks as shown in Figure~\ref{fig:dgx1-topo}.  


The \dgxone's design was heavily influenced by the need to do gradient reduction for machine learning workloads. Specifically, this topology forms two non-overlapping rings: one connecting nodes $\{0,1,4,5,6,7,2,3\}$ with two NVLinks per edge and another connecting $\{0,2,1,3,6,4,7,5\}$ with one NVLink per edge. These rings are bidirectional and thus form $6$ logical single-NVLink rings. The NCCL library implements \allgather by running $6$ simultaneous ring algorithms as we discuss below.  


\subsection{Cost Model}
  
We will characterize the communication cost using the $(\alpha, \beta)$ model~\cite{hockney1994communication}. That is, sending a message of size $L$ along a link costs $\alpha + L\cdot\beta$ time. 
Here, $\alpha$ is the latency of communication and captures the {\em fixed} costs, such as the overhead of initiating a transfer or invoking a GPU kernel, 
and $\beta$ is the inverse bandwidth of the link and captures {\em per-byte} costs, such as copying data into system buffers. Li \etal{} extensively studies the transfer time of buffers with 
different sizes over numerous GPU interconnections\cite{alphabeta}. Their result show that with NVLinks, the transfer time stays almost constant up-to a large buffer size and only then it start to increase linearly. 
These results confirm that the $(\alpha,\beta)$ model is suitable for characterizing communication cost over NVLinks.

The cost of a collective algorithm for an input of size $L$ will be of the form $a\cdot\alpha + b \cdot L \cdot \beta$. We call $a$ the {\em latency cost} of the algorithm and $b$ the {\em bandwidth cost} of the algorithm. Given a class of algorithms that implement a collective on a given topology, an algorithm is {\em latency-optimal} ({\em bandwidth-optimal}) if no other algorithm in the class has a lower latency (bandwidth) cost. Usually, there is a tradeoff between the latency cost and the bandwidth cost when designing collective algorithms.  An algorithm with latency cost $a$ and bandwidth cost $b$ is said to be {\em Pareto-optimal} with respect to a class of algorithms if for every algorithm in the class with latency cost $a'$ and bandwidth cost $b'$, we have $a = a' \Rightarrow b' \geq b$ and $b = b' \Rightarrow  a' \geq a$.


\subsection{Bandwidth-Optimal Algorithm for \dgxone}
\label{sec:motivation:bw-optimal}
As described above, the \dgxone topology has $6$ logical rings. \allgather for one ring can be implemented as follows. Each node simultaneously sends its data to the next node in the ring. In subsequent steps, each node stores the received data and sends it to the next node in the ring. In $7$ steps all nodes will have received data from all of the other $7$ GPUs. The $6$-ring algorithm is a generalization of this algorithm. Each node splits its data into $6$ chunks and executes the ring algorithm along each of the $6$ rings, with one chunk per ring. If $L$ is the size of the input data, each ring algorithm takes $7$ steps and communicates $\frac{L}{6}$ bytes. Thus, the cost of the $6$-ring algorithm is  
$$7\cdot \alpha + \frac{7}{6}\cdot L \cdot \beta$$

Each node has to receive at least $7 \cdot L$ amount of data, and it has an agglomerated incoming per-byte cost of $\beta/6$ (6 incoming NVLinks). Thus, any algorithm for \allgather has to take at least $\frac{7}{6}\cdot L \cdot \beta$ amount of time. Thus, this algorithm is bandwidth-optimal for the \dgxone topology. But can we do better with the latency cost? 

Using the techniques described in this paper, we have automatically synthesized an algorithm~(Section~\ref{fig:dgxone:syn}) with cost $$3\cdot \alpha + \frac{7}{6}\cdot L \cdot \beta$$ To the best of our knowledge, this algorithm was not previously known. Moreover, we prove that this algorithm is Pareto-optimal with respect to the class of algorithms we call $k$-synchronous algorithms~(Section~\ref{sec:ksync}).  

\subsection{Latency-Optimal Algorithm for \dgxone}
The next question is whether we can improve upon the latency cost of the synthesized algorithm. If each node communicates its data along a binary tree instead of a ring, it would take at least $3$ steps. Using the techniques described in this paper, we have automatically synthesized a better algorithm~(Section~\ref{fig:dgxone:syn}) with cost $$2\cdot \alpha + \frac{3}{2}\cdot L \cdot \beta$$
Since the \dgxone topology has a diameter of $2$, this algorithm is latency-optimal. To the best of our knowledge, a latency-optimal algorithm for the \dgxone was not previously known. This algorithm is Pareto-optimal with respect to the class of $k$-synchronous algorithms.

\section{Algorithm Synthesis}
\label{sec:synthesis}
This section demonstrates a method to synthesize Pareto-optimal algorithms that implement a collective primitive on a given topology. The Pareto-optimality is defined with respect to a class of algorithms we call {\em $k$-synchronous} algorithms.

We distinguish between {\em \reducing} collectives such as \allreduce and \reducescatter that combine chunks through computation, and {\em \broadcasting} collectives such as \allgather and \broadcast that simply transfer data among nodes. We will focus on synthesizing \broadcasting collectives and show how to derive \reducing collectives from related \broadcasting ones. 

\subsection{$k$-synchronous Algorithms}
\label{sec:ksync}
\begin{figure} 
    \includegraphics[width=\columnwidth]{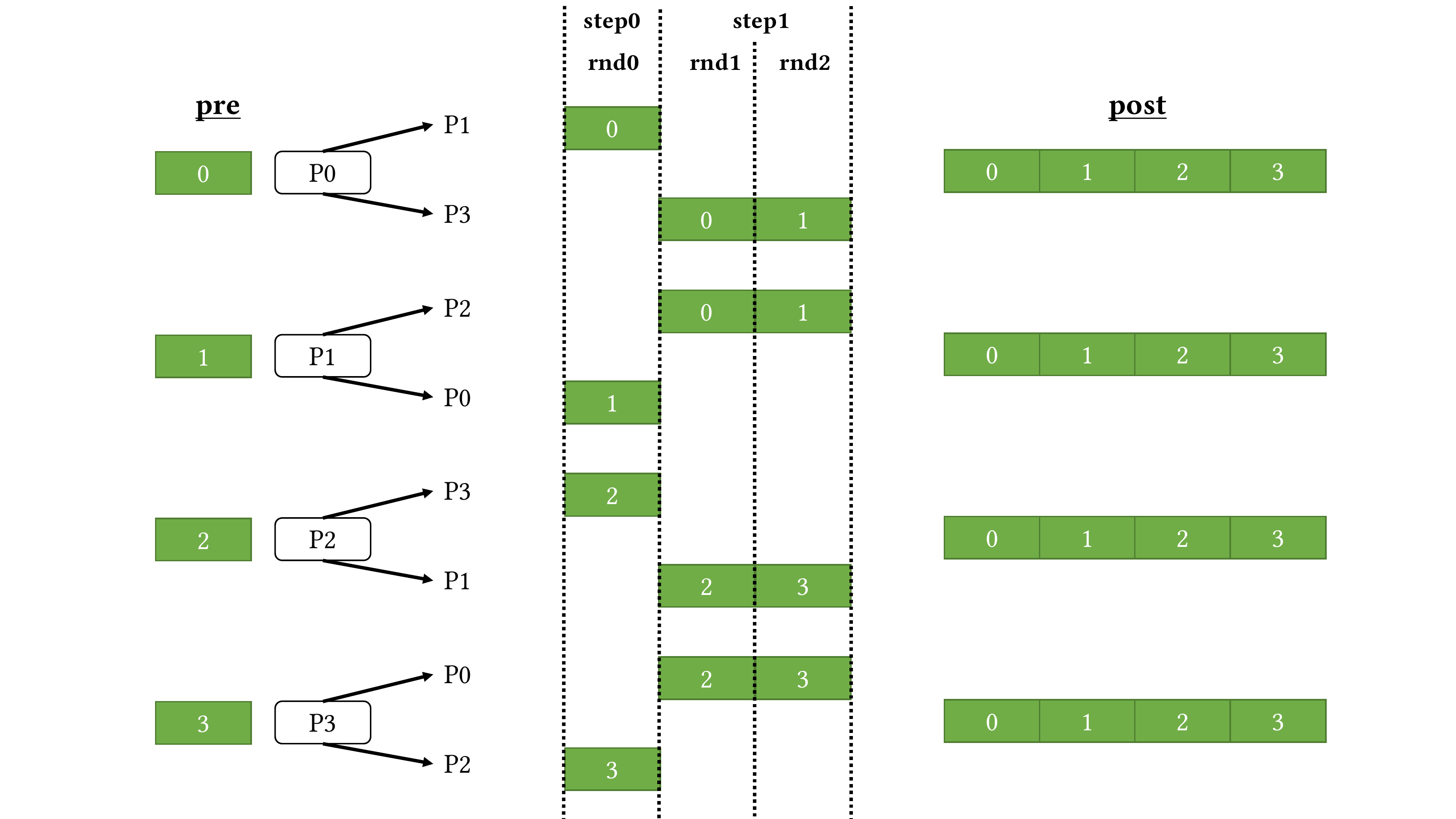}
    \caption{A $1$-synchronous algorithm for \allgather on a ring topology.}
    \label{fig:allgatherex}
\end{figure}
Figure~\ref{fig:allgatherex} shows the recursive-doubling~\cite{thakur2005optimization} algorithm for \allgather for a ring topology of four nodes $P0, P1, P2, P3$ with four bidirectional links of equal bandwidth. This algorithm proceeds in two {\em steps}. In the first step, nodes at "distance" 1, namely $P0, P1$ and $P2, P3$ send their data to each other. Each node now has data from two nodes, which it communicates entirely with nodes at distance 2, i.e., nodes $P0, P3$ and $P1, P2$ in the second step. At the end, each node has data from every other node. Since the second step involves sending twice the amount of data as the first step, we say it has two \emph{rounds} where in each round, it sends data. Thus, this step has a total of $3$ rounds. Of the eight (unidirectional) links, this algorithm uses only four of them per step. To improve bandwidth utilization, a better option is to split the input data into equal-sized {\em chunks} and communicate them independently. For instance, the ring algorithm described in Section~\ref{sec:motivation:bw-optimal} uses $3$ chunks per node. 

The algorithm in Figure~\ref{fig:allgatherex} and many classical collective algorithms~\cite{thakur2005optimization,chan2007collective} are instances of {\em synchronous} algorithms. A synchronous algorithm proceeds in a sequence of synchronous communication {\em steps} with nodes waiting for other nodes to finish their rounds before starting the next step. Even if an implementation might not enforce a global barrier across the nodes, these algorithms choose the amount of data to communicate per step based on the bandwidth constraints so that the nodes finish each step at (roughly) the same time.  

Many algorithms, like the one in Figure~\ref{fig:allgatherex}, communicate different numbers of chunks per step. We consider each step as consisting of multiple rounds with each node sending at most one chunk per unit-bandwidth on its outgoing links. Intuitively, the number of rounds in an algorithm controls its bandwidth cost, while the number of steps controls its latency cost. A synchronous algorithm with $\steps$ steps and $\rounds$ rounds is {\em $k$-synchronous} if $\rounds \leq \steps + k$. The parameter $k$ limits the amount of communication per step and allows an SMT solver to effectively search the space of algorithms bounded by that $k$.  


\subsection{\broadcastingCap Collective Instance}
Now we will provide a uniform formulation for representing $k$-synchronous algorithms for \broadcasting collectives. 
An instance of \collectiveproblem is a tuple $(\gchunk,\steps,\rounds,\size,\bw,\pre,\post)$, where
\begin{itemize}
    \item[] \hspace{-0.5cm}Parameters:
    \begin{itemize} 
    \item $\gchunk\in\posint$ is the global number of chunks
    \item $\steps\in\posint$ is the total number of steps
    \item $\rounds\in\posint$ is the total number of rounds
    \end{itemize}
    \item[] \hspace{-0.5cm}Topology:
    \begin{itemize} 
    \item $\size\in\posint$ is the number of nodes
    \item $\bw\subseteq\powerset(\range{\size}\times\range{\size})\times\mathbb{N}$ is the bandwidth relation
    \end{itemize}
    \item[] \hspace{-0.5cm} Specification:
    \begin{itemize} 
        \item $\pre\subseteq\range{\gchunk}\times\range{\size}$ is the pre-condition
        \item $\post\subseteq\range{\gchunk}\times\range{\size}$ is the post-condition
    \end{itemize}
\end{itemize}
Note that for a set $M$ we write $\powerset(M)$ for the power set of $M$, i.e., the set of all subsets. For an integer $x$, we write $\range{x}$ for the set $\{0, 1, \ldots, x\}$. 
Here, $\gchunk, \steps, \rounds$ are parameters to the desired $k$-synchronous algorithm. The rest are explained below. 

\subsubsection{Topology} 
\label{sec:topology}
$\size$ is the number of nodes in the topology. $\bw$ gives a flexible way to express different bandwidth constraints we have seen in practice. In its most general form, $\bw$ bounds the sum of chunks sent along a set of edges in a single round. A point-to-point communication link from $s$ to $d$ with maximum bandwidth (in chunks per round) $b$ can be modeled by $(\{(s, d)\}, b) \in \bw$. Some topologies might limit the net outgoing bandwidth $b$ from a certain node $s$. If $E$ is the set of outgoing neighbors of $s$, we can model this by $(\{(s, e) \mid e \in E\}, b) \in \bw$. To model shared bus topologies, where only one node can send in a round, we include $(\{(a, b) \mid a \in N, b \in N\}, b)$ in $\bw$ for the set of nodes $N$ sharing the same link. Note that these constraints are per round, and when performing $r_i$ rounds in step $i$, we simply multiply the bandwidth constraint by $r_i$.

\newcommand{\relAll}{All\xspace}
\newcommand{\relRoot}{Root\xspace}
\newcommand{\relScattered}{Scattered\xspace}
\newcommand{\relTranspose}{Transpose\xspace}
\newcommand{\chunkReduce}{\left\lfloor\frac{i}{\size}\right\rfloor}
\begin{table}
    \begin{tabularx}{\columnwidth}{@{}Xl@{}}
        \toprule
        Name & Relation \\
        \midrule
        \relAll & $\range{\gchunk}\times\range{\size}$ \\
        \relRoot & $\range{\gchunk}\times\{n_\mathit{root}\}$ \\
        \relScattered & $\{(c,n)\in\range{\gchunk}\times\range{\size} \setwhere n=c\bmod \size\}$ \\
        \relTranspose & $\{(c,n)\in\range{\gchunk}\times\range{\size} \setwhere n=\left\lfloor\frac{c}{\size}\right\rfloor\bmod \size\}$ \\
        \bottomrule
    \end{tabularx}
    \caption{Common relations in pre- and post-conditions of collective primitives.}
    \label{tbl:relations}
\end{table}
\begin{table}
    \begin{tabularx}{\columnwidth}{@{}Xll@{}}
        \toprule
        Collective & $\pre$ & $\post$ \\
        \midrule
        \gathercoll & \relScattered & \relRoot  \\
        \allgather & \relScattered & \relAll  \\
        \alltoall & \relScattered & \relTranspose  \\
        \broadcast & \relRoot & \relAll \\
        \scatter & \relRoot & \relScattered  \\
        \bottomrule
    \end{tabularx}
    \caption{Specifications of collective primitives as \collectiveproblem instances using a small set of common relations for pre- and post-conditions.}
    \label{tbl:collectives}
\end{table}

\subsubsection{Collective Specification}
\label{sec:specifications}
The $\pre$ relation specifies the nodes where the chunks reside at the beginning of the algorithm and the $\post$ relation specifies the set of nodes where a chunk needs to be transferred to. Table~\ref{tbl:relations} specifies useful relations that can be used to specify common collectives as shown in Table~\ref{tbl:collectives}. For instance, \allgather starts in a state where chunks are in the \relScattered relation in Table~\ref{tbl:relations}. In other words, the $c$ chunks of the input at node $n$ are given chunk identifier $i\cdot P + n$ for $0 \leq i < c$. From this \relScattered state, \allgather requires all the input chunks to be copied to all nodes, as specified by \relAll relation in Table~\ref{tbl:relations}. Similarly, \broadcast requires all the chunks from the root $n_{root}$ to be copied to all nodes.

While \collectiveproblem uses a global number of chunks $\gchunk$, it is more
typical in existing literature to consider the per-node number of chunks
$\chunk$. We will use the per-node number when discussing the cost model and
search algorithm in Sections~\ref{sec:costmodel} and \ref{sec:pareto:optimal}
and when presenting our evaluation in Section~\ref{sec:evaluation}. Note that
how these two counts relate to each other is collective dependent: for
\broadcast $\gchunk=\chunk$, while for \allgather $\gchunk=\size\cdot\chunk$.
The formalization must still use a global numbering of chunks, as some exotic
collectives, e.g. MPI's Allgatherv, may not have a single per-node chunk count.

\subsection{Candidate Solution}
Given an instance of \collectiveproblem $(\gchunk,\steps,\rounds,\size,\bw,\pre,\post)$, a candidate solution is a pair $(\rparts,\sends)$. Here $\rparts$ is a sequence $r_0,\allowbreak r_1,\allowbreak \dotsc,\allowbreak r_{\steps -1}$ such that $\sum_i r_i = \rounds$ and denotes the number of rounds per step. $\sends$ is a set of sends of the form $(c,n,n',s)$, which specifies that chunk $c$ must be sent from node $n$ to node $n'$ at step $s$. This defines a {\em run} defined as a  sequence $V_0,V_1,\dotsc,V_{\steps}$ such that $V_0 = \pre$ and for all $0 \leq s < \steps$, $V_{s+1}$ reflects the chunks present at a given node after accounting for the sends at step $s$:
$$ V_{s+1} = V_{s} \cup \{(c,n') \mid (c, n) \in V_{s} \wedge (c, n, n', s) \in \sends\} $$ 

This candidate solution is a valid $k$-synchronous algorithm for the instance if $V_{\steps} \subseteq \post$ and the following bandwidth constraint hold
\begin{align*}
    &\begin{aligned}
        \forall &s\in\range{\steps},\,(L,b)\in\bw \qst \\
        & |\{(c,n,n',s)\in \sends \setwhere (n,n')\in L\}|\leq b \cdot r_s
    \end{aligned} 
\end{align*}
At each step $s$ consisting of $r_s$ rounds, the number of sends in each link should be bounded by the bandwidth constraint multiplied by $r_s$.  

\subsection{SMT Encoding for \broadcastingCap Collectives}
\label{sec:encoding}
Given an instance, the SMT encoding incorporates the constraints above allowing the SMT solver to systematically search over candidate solutions $(\rparts, \sends)$. It is straightforward to encode each $r_s$ of $\rparts$ as integer variables whose sum is $\rounds$. In contrast, one has to be careful in encoding $\sends$. For instance, our initial attempt to encode every tuple $(c, n, n', s) \in \sends$ as a Boolean variable was not successful, because Z3, the SMT solver we used, did not solve larger problem instances fast enough. One way we were able to scale Z3 is to use a careful combination of Boolean, integer, and pseudo-Boolean constraints as we describe below.

We split the encoding of $\sends$ into integer variables $\start{c}{n} \geq 0$, indicating the earliest step a chunk $c$ becomes available at node $n$ and Boolean variables $\send{n}{c}{n'}$ determining whether a node $n$ sends chunk $c$ to $n'$ (at any step). 
\newcommand{\edges}{E}
To help with pruning the encoding, let $\edges=\{(n,n') \setwhere \allowbreak \forall (L,b)\in\bw \qst (n,n')\in L \Rightarrow b > 0\}$, i.e., the pairs of nodes with non-zero bandwidth between them. Pseudo-Boolean constraints allow one to use Boolean variables as $0,1$ integers which we will use in the exposition below. 

The following two constraints enforce the pre- and post-conditions
\begin{align}
    \forall (c,n) \in \pre \ \ \start{c}{n} &=0
    \tag{C1}\label{eqn:bc-zero} \\
    \forall (c,n) \in \post \ \ \start{c}{n}&\leq \steps
    \tag{C2}\label{eqn:bc-insteps}
\end{align}
If a chunk becomes available in a node, but is not part of the precondition, then the node should have received the chunk from some other node. For optimality, we also enforce that the node does not redundantly receive the chunk more than once.  
\begin{align}
    \forall (c,n) \not\in \pre \ \ \start{c}{n}&\leq \steps \Rightarrow \Sigma_{(n',n)\in\edges}\,\send{n'}{c}{n}=1
    \tag{C3}\label{eqn:bc-hassender}
\end{align}
To send a chunk, it must exist on the source node before it is received on the destination node. 
\begin{align}
    \forall (c,n) \in E \ \ \send{n}{c}{n'}\Rightarrow\start{c}{n}<\start{c}{n'}
    \tag{C4}\label{eqn:bc-sendexisting}
\end{align}
The following enforces the bandwidth constraint at all steps $1 \leq s \leq \steps$ and bandwidth constraint $(L,b)\in\bw$:
\begin{align}
    \Sigma_{(c,(n,n'))\in\range{\gchunk}\times L}\left(\send{n}{c}{n'}\wedge\start{c}{n'}=s\right)\leq b \cdot r_s
    \tag{C5}\label{eqn:bc-bw}
\end{align}
Note, we have multiplied the bandwidth constraints by $r_s$ to allow $r_s$ rounds at step $s$. 
Finally, the following bounds the total rounds $\rounds$:
\begin{align}
    \Sigma_{1\leq s\leq\steps}(r_s)=\rounds
    \tag{C6}\label{eqn:bc-rounds}
\end{align}

Once the problem instance has been encoded, the SMT solver will attempt to find a model $M$, which maps the variables $\start{c}{n}$, $\send{n}{c}{n'}$ and $r_s$ to concrete values such that Constraints \ref{eqn:bc-zero} through \ref{eqn:bc-rounds} are satisfied. If a model exists then an algorithm $(\rparts, \sends)$ can be constructed with:
\begin{align*}
    \rparts&=M(r_0),\dotsc,M(r_{\steps-1}) \\
    \sends&=\{(c,n,n',t) \setwhere M(\send{n}{c}{n'}) \wedge M(\start{c}{n}) = t+1 \}
\end{align*}
If the SMT solver says the problem is unsatisfiable, then no algorithm exists for the problem instance.

\subsection{\reducingCap Collectives}
\label{sec:reduction}
It is well known that certain \reducing collectives are {\em inverses} of \broadcasting collectives. For instance, a \reduce algorithm can be generated by inverting an algorithm for \broadcast on a topology where all links have been reversed. Intuitively, whenever the \broadcast sends the same chunk to two different nodes, in its inverse the \reduce algorithm will receive the two {\em versions} of the chunk from these nodes and apply the reduction operation. The node will send the resulting chunk to the node it received the chunk from in the \broadcast. Similarly, we can generate \reducescatter algorithms by inverting \allgather algorithms. 

Generally the inverting procedure works for any \reducing collective that has a single root node for each chunk. Notably, this does not include \allreduce, which replicates the result onto all nodes. For synthesizing \allreduce algorithms, we first notice that \allreduce can be expressed as a combination of \reducescatter followed by an \allgather. We synthesize \allreduce algorithms by synthesizing an \allgather algorithm and preceding it with its inverse \reducescatter algorithm.

\subsection{Cost Model}
\label{sec:costmodel}
Say we have synthesized a $k$-synchronous algorithm with $\chunk$ chunks, $\steps$ steps, and $\rounds$ rounds. 
We will use the $(\alpha, \beta)$ cost model~\cite{hockney1994communication} to evaluate cost of this algorithm. 
Here, $\alpha$ is the latency of each link in the topology and $\beta$ is the time taken sending a byte along a unit-bandwidth link. 
If the input data of $L$ bytes is divided into $\chunk$ chunks, a step $s$ with $r_s$ rounds takes $\alpha + \frac{r_s}{\chunk}\cdot L \cdot \beta$ time. Therefore, the entire algorithm will finish in time
$$ \steps \cdot \alpha + \frac{\rounds}{\chunk} \cdot L \cdot \beta $$

\subsection{Pareto-optimal Algorithms}
\label{sec:pareto:optimal}
The discussion above shows that for a given topology and a collective with an input size $L$, the cost of a $k$-synchronous algorithm can be characterized by the tuple $(\steps, \frac{\rounds}{\chunk})$. An algorithm with cost $(a,b)$ is {\em Pareto-optimal} with respect to the class of $k$-synchronous algorithms if for every algorithm in this class with cost  ($a', b')$ we have $a = a' \Rightarrow b' \geq b$ and $b = b' \Rightarrow a' \geq a$. 
An algorithm with cost $(a,b)$ is considered {\em latency-optimal} ({\em bandwidth-optimal}), if for every $k$-synchronous algorithm with cost 
$(a',b')$ we have $a' \geq a$ ($b' \geq b$). 

Note that latency- or bandwidth-optimal algorithms are not necessarily Pareto-optimal as they can be "wasteful" in the other parameter. Pareto-optimal algorithms form a {\em Pareto-frontier} with different algorithms in the frontier being better than others for a given input size $L$ based on the $\alpha$ and $\beta$ parameters of the topology. 

\begin{algorithm}
	\caption{Synthesizing Pareto-Optimal Algorithms} 
    \begin{algorithmic}[1]    
        \Procedure{Pareto-Synthesize}{$k, \mathit{Coll}, \size, \bw$}
        \State $a_l = \mathit{Diameter(\size, \bw)}$
        \State $b_l = \mathit{InvBisectionBandwidth(\size, \bw)}$
        \State $(\pre, \post) = \mathit{Lookup(Coll)}$ \Comment{Table~\ref{tbl:collectives}}
        \For {$\steps=a_l,a_l+1\ldots$}
            \State $A = \{(\rounds,\chunk) \mid \steps \leq \rounds \leq \steps+k \wedge \frac{\rounds}{\chunk} \geq b_l\}$
            \For {$(R,\chunk) \in A$ in ascending order of $\frac{R}{\chunk}$}
                \State $\gchunk=\toglobal(\mathit{Coll},\chunk)$
                \If {$\mathit{SMT(\gchunk, \steps, \rounds, \size, \bw, \pre, \post) = SAT}$}
                \State Report synthesized algorithm $(\steps,\rounds,\chunk)$
                \If {$\frac{\rounds}{\chunk} = b_l$}
                \State {\bf return}
                \EndIf
                \State {\bf break}
                \EndIf
			\EndFor
        \EndFor
        \EndProcedure
	\end{algorithmic} 
\end{algorithm}

The procedure above systematically synthesizes Pareto-optimal $k$-synchronous algorithms. The inputs are the parameter $k$, the name of the collective to synthesize, and the topology parameters $\size, \bw$~(Section~\ref{sec:topology}). The procedure computes the latency lower bound $a_l$ from the diameter of the topology, and the bandwidth lower bound $b_l$ from the inverse bisectional bandwidth of the topology. The procedure starts enumerating steps $\steps$ starting with $a_l$. Then it generates $A$, the candidate set of tuples $(\rounds, \chunk)$ that satisfy the round constraint and the inverse bandwidth constraint. Note that without the $k$ parameter, this set would be unbounded. The procedure checks if a $(\steps,\rounds,\chunk)$ algorithm exists in the increasing order of the bandwidth cost $\frac{\rounds}{\chunk}$ using the encoding discussed in Section~\ref{sec:encoding}. If one exists, the reported algorithm is guaranteed to be Pareto-optimal for the current steps $\steps$. As we increase the number of $\steps$, we get algorithms with lower bandwidth cost. Additionally, if the current bandwidth cost matches the lower bound $b_l$, the procedure returns. As we have already generated the Pareto-optimal algorithm with $b_l$ bandwidth cost, it is not necessary to increase $\steps$ further. 
Note, that it is possible for this procedure to never terminate as there can sometimes be unbounded number of Pareto-optimal algorithms for certain topologies and collectives. While the synthesis procedure above is for \broadcasting collectives, synthesis for \reducing collectives is similar~(Section~\ref{sec:reduction}). 

\section{Code Generation}
\label{sec:lowering}
The prior section described a synthesis procedure for generating Pareto-optimal algorithms.
This section describes a tool called \tool{} that implements this procedure and generates 
high-performance collective implementations for both NVIDIA and AMD GPUs.

Every synthesized algorithm, at its core, is a sequence of commands that
describe \emph{what} data needs to be sent (i.e., which chunk),
\emph{where} it needs to be sent (i.e., a source and destination),
\emph{when} it needs to be sent (i.e., during which synchronous step), and \emph{with} which chunk(s) it needs to be reduced. 
\tool{} generates SPMD multi-process \CC{} code combined with CUDA kernels that implement these commands. 

Each GPU involved in the computation has its own code as part of a top-level switch statement. Communication between GPUs is enabled using CUDA IPC
memory handles, which allows a GPU to access a remote GPU's memory using shared pointers. Thus, communication between GPUs simply involves 
writing data to appropriate buffers. However, there are a few crucial choices that impact the communication performance.   





\paragraph{DMA engines and kernel copies:} Data may be moved either by executing load or store instructions through a kernel, or by using a specialized DMA engine via \texttt{cudaMemcpy}. A kernel copy allows data movement and computation to be fused in a kernel while a DMA engine has a higher initial $\alpha$ cost but may have higher bandwidth, leading to a lower $\beta$ cost. On NVLink, DMA engine bandwidth is about 10\% better than kernel copy bandwidth, due to details of the wire-level protocol. Transfers are packetized, with each packet including a header (containing address, error correction data, etc.) and a variable-length payload. DMA engines are able to emit maximum-sized packets, but kernel copy packets are limited to the 128-byte cache line size.

\paragraph{Push and pull models:} Each DMA engine is located on a particular GPU. Data movement between two GPUs can be executed by either the receiver's DMA engine (a {\em pull} model) or by the sender's DMA engine (a {\em push} model). Kernel copies have the same two approaches. This may have performance implications due to the link protocol: the push model only needs to send write request packets with a payload, whereas a pull model first sends request packets and then receives response packets with data. When communicating bidirectionally, the request packets reduce the bandwidth available for the response packets. Thus, even though the push model may require extra memory, we have found it to be up to 10\% faster than the pull model.



\paragraph{Single and multiple kernels:} One way to implement a synthesized algorithm is by emitting several kernels, one per step, which forces a global synchronization between steps and, as a consequence, introduces large overheads.
Alternatively, \tool{} fuses all steps into one kernel and thus we implement the synchronizations between GPUs as a fine-grained signal and wait mechanism with shared flags. In our single kernel implementation, each chunk for each connection has a dedicated flag; a chunk on a GPU is valid 
only when the associated flag is set. There is a \texttt{\_\_threadfence\_system()} between the data movement operations and the operation to set the flag on the remote GPU signals that the transfer is complete. 

\paragraph{Size and Number of Thread Blocks:} \tool{} dedicates a given number of thread blocks to each link and for each step, it uses the same number of thread blocks to communicate through that link. 
For different input sizes, the number of thread blocks significantly affects performance and in later sections we show how we empirically search for the fastest configuration for various input sizes.

\section{Evaluation}
\label{sec:evaluation}


This section demonstrates how we model and synthesize collectives for
two multi-GPU systems with proprietary interconnects used for training large
deep learning models. In both cases, we
demonstrate 1) how to model the interconnect using \tool, 2) what
transport we utilize in lowering synthesized collectives, and 3) the
Pareto-frontier of algorithms we find for the respective
interconnects.

\subsection{Hardware}
The following section describes the hardware topology we model for our
NVIDIA and AMD machines.

\subsubsection{NVIDIA DGX-1: 8 V100 GPUs}
A DGX-1 is a multi-GPU server sold directly by NVIDIA in addition to
being a pay-as-you-go rental option in most cloud providers.  It
contains two 20-core Intel Xeon E5-2698 v4 processors with 512 GB DRAM
split across the two sockets, along with 8 NVIDIA V100 GPUs, each
with 32 GB of HBM2 memory. The GPUs are connected using NVIDIA's proprietary NVLink
interconnect; each GPU has 6 25 GB/s NVLink ports.  Figure~\ref{fig:dgx1-topo} shows the
topology: the 8 GPUs are interconnected with 2 non-overlapping
Hamiltonian cycles.  One of those cycles has two
NVLink connections between each pair of GPUs. The GPUs are also connected to the CPUs by PCIe 3.0 x16 links, but we do not use them
due to the wide disparity between per-GPU NVLink and PCIe bandwidth ($\sim$150~GB/s vs. $\sim$14~GB/s).
We also run synthesis on this platform.

\begin{figure}
\includegraphics[page=2,width=\columnwidth]{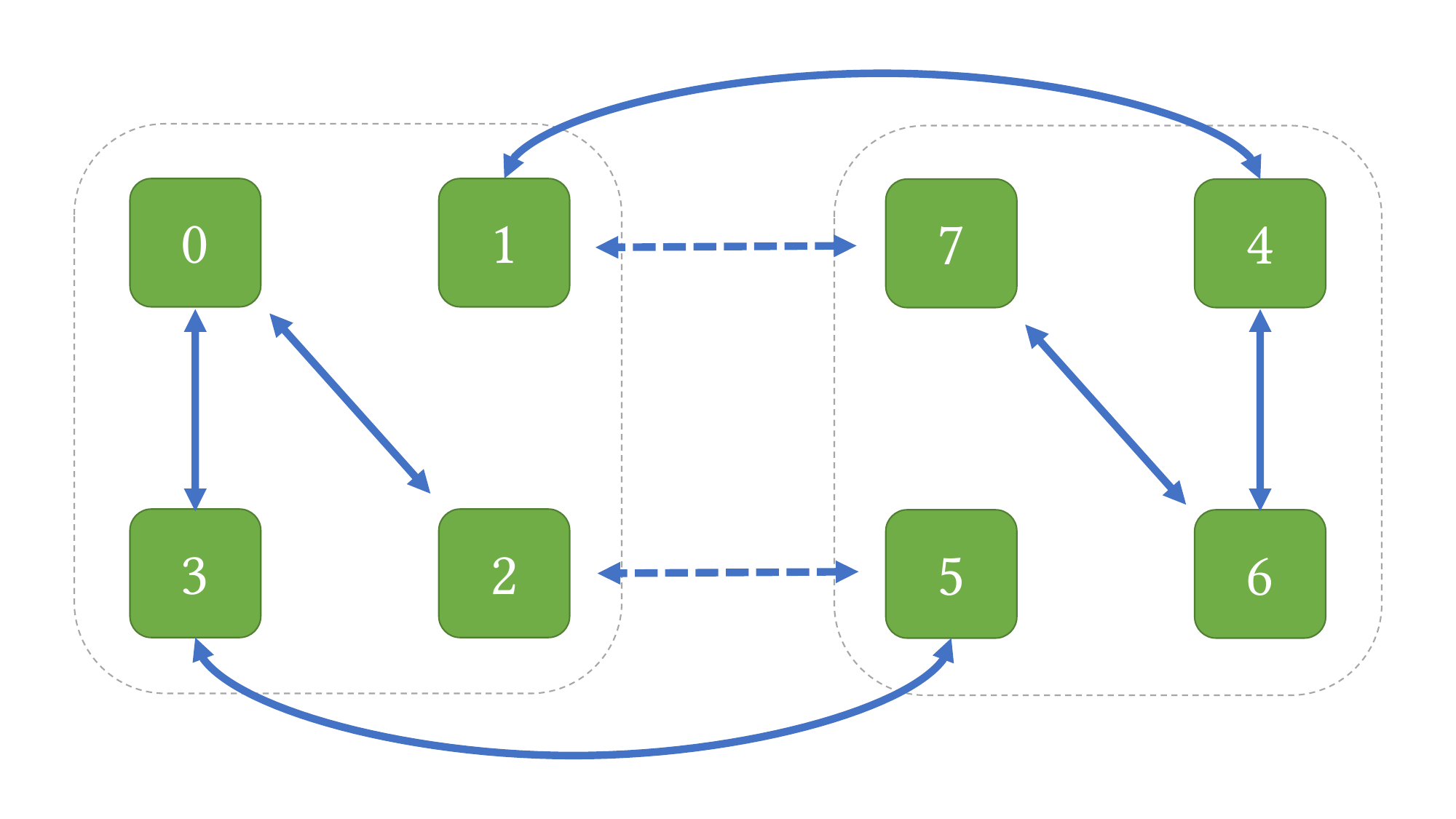}
\caption{Topology of a Gigabyte MI50 8 GPU AMD System.}
\label{fig:amd-topo}
\end{figure}

\subsubsection{Gigabyte Z52: 8 AMD MI50 GPUs}
A Gigabyte Z52 system is a consumer grade multi-GPU system.
It has two 64-core AMD EPYC 7002 processors with 1 TB DRAM split across the two sockets, as well as 
8 AMD MI50 GPUs, each with 32 GB of HBM2 memory. 4 GPUs are connected to each socket with PCIe links, denoted by a box in Figure~\ref{fig:amd-topo}.
Like NVIDIA, AMD also provides a proprietary high-speed
interconnect called xGMI that links GPUs together.  Each blue line is
an xGMI link between a pair of GPUs.  Note that the xGMI connections
build two disconnected islands: 3 GPUs per island are on 1 socket
while a lone GPU is on the \emph{other} socket (i.e., GPU 1 and 5).
The Gigabyte system uses PCIe 4.0 x16 links with measured bandwidth ($\sim$27 GB/s)
that approaches xGMI's measured bandwidth ($\sim$33 GB/s). As such, we use PCIe to
connect the rings.

\subsection{Modeling Bandwidth Constraints}
The hardware in this paper has distinct and interesting
topologies. This section describes how we model those respective
topologies in \tool.

\subsubsection{NVIDIA DGX-1: 8 V100 GPUs}
Each NVLink connection is point-to-point; thus our bandwidth constraints are simply
the enumeration of each pair of GPUs connected via NVLink. As each NVLink
connection can send 1 chunk per round, $\bw$ has entries $(\{(n,n')\},1)$ for each
pair of GPUs in one cycle and entries $(\{(n,n')\},2)$ for GPUs in the other.

\subsubsection{Gigabyte Z52: 8 AMD MI50 GPUs}
Unlike NVLink, xGMI connections are not simply point-to-point but also
transparently act as a router.  For example, GPU 2 can send a message
to GPU 3 even though they lack a physical connection: GPU 0 routes
messages on GPU 2's behalf.  However, this utilizes multiple links,
and thus if GPU 0 concurrently sends a message to GPU 3, it can expect
half the bandwidth of the link.  We thus only model the direct
connections in Figure~\ref{fig:amd-topo}.  One way to connect the
rings is to utilize PCIe and let GPU 1 connect to all other GPUs
within its same socket (0, 2, and 3) and GPU 5 connect to GPUs within
its same socket (4, 6, and 7).  Because PCIe is shared, we could also
enforce that only 1 PCIe connection occurs on every round, per socket.
%
For example, the entry in $\bw$ for the left socket is $(\{(0,1),(1,0),(1,2),(2,1),(1,3),(3,1)\},1)$.
However, we were unable to utilize both xGMI and PCIe at the same time
so our model of the bandwidth ignores the dotted xGMI connections in
Figure~\ref{fig:amd-topo}.  As such, we explicitly model the topology as a ring with GPUs 1 and 5 connecting the xGMI islands. Lastly, because the bisection bandwidth between the two xGMI islands is limited by the
PCIe links that connect them, any bandwidth optimal algorithm will be limited by the bandwidth of these PCIe links. Therefore, we model the same $\beta$ cost
for xGMI and PCIe and assume all
links can send a single chunk per step.

\subsection{NCCL and RCCL Baselines}
We use NCCL (version 2.7.8-1) and RCCL (installed from ROCm 3.5.0) for
baselines on NVIDIA and AMD hardware, respectively.  NCCL is a
hand-written and optimized communication library from NVIDIA. RCCL is
a port of NCCL that uses the ROCm HIP compiler and targets AMD
hardware. They share the same core algorithms and differ only in how
they interact with the underlying hardware.

\newcolumntype{H}{>{\setbox0=\hbox\bgroup}c<{\egroup}@{}}

\begin{table}
  \begin{tabular}{@{}llll@{}}
\toprule
Collective &$\chunk$ & $\steps$ & $\rounds$ \\
\midrule
\allgather/\reducescatter & 6 & 7 & 7 \\ 
\allreduce & 48 & 14 & 14 \\ 
\broadcast/\reduce  & $6m$ & $6+m$ & $6+m$ \\ 
\bottomrule
\end{tabular}
\caption{NCCL hand-written collectives and their chunks and steps. For \reducescatter $\chunk$ should be multiplied by 8.}
\label{table:nccl}
\end{table}

Table~\ref{table:nccl} gives an overview of the collectives that NCCL
implements and number of chunks and steps they use on a \dgxone. NCCL's
algorithms are all based on either rings or trees. However,
Table~\ref{table:nccl} uses only ring algorithms, as we observed that on
DGX-1 NCCL's trees are just simple paths, which are no better than
using rings for any input size.

Our analysis of the chunks ($\chunk$), steps ($\steps$), and rounds($\rounds$) is from our manual
inspection of the NCCL source. For \reduce and \broadcast NCCL implements a
pipelined algorithm, which chooses a multiplier $m$ such that chunks stay
approximately equally sized. Their running times are then $(6+m) \cdot \alpha + \frac{6+m}{6m}\cdot L \cdot \beta$
and they get closer to bandwidth optimality as $m$ gets larger.

As we show in the next section, \tool{} is able to synthesize all
these NCCL collectives and more, including \scatter, \gathercoll, and
\alltoall.  


\subsection{Synthesizing Collective Algorithms}
Table~\ref{fig:dgxone:syn} and Table~\ref{fig:amd:syn} enumerate
various algorithms we synthesize for NVIDIA \dgxone and Gigabyte's
\amd architecture.  For each collective, we synthesize a latency and
bandwidth optimal implementation, along with others that exist at
various points along the latency-bandwidth curve. The first column
combines collectives which are the inverse of each other (i.e.,
\scatter and \gathercoll) and those that can be reduced to the 
\broadcasting collective using the reduction explained in
Section~\ref{sec:reduction} (e.g. \reduce to \broadcast).

\subsubsection{Optimality}
Note we find many latency and bandwidth optimal algorithms
for each collective, as we search over $k$-synchronous algorithms for different values of $k$. 
Consider the \allgather collective: we find many algorithms with various numbers
of steps. However, the latency optimal algorithms (2 steps) dominate all others
in the $\alpha$ term of the cost model.  Likewise, the bandwidth optimal
algorithms dominate all others with their low ratio of rounds to chunks ($7/6$).
We synthesized algorithms in the $0$-synchronous class ($\rounds = \chunk$) as 
the code generation is much easier. 


Note that NCCL's \allgather algorithm is bandwidth optimal, 
and while it is also the lowest latency algorithm that NCCL provides, it is not latency optimal. We
are able to synthesize both a bandwidth optimal algorithm with better latency
(6-chunks 3-steps 7-rounds), as well as a latency optimal algorithm.
In general, our
synthesized latency optimal algorithms have no counterpart in NCCL
and our bandwidth optimal algorithms are better than NCCL's for \allgather,
\broadcast, and \reduce.


\begin{table}
  \begin{tabularx}{\columnwidth}{@{}lHlllXHr@{}}
\toprule
Collective & Topology & $\chunk$ & $\steps$ & $\rounds$ & Optimality & Running Time & Time \\
\midrule
\allgather & \dgxone & 1 & 2 & 2 &Latency&$2 \cdot \alpha + 2\cdot L \cdot \beta$& 0.3 s\\
(\reducescatter)& \dgxone & 2 & 3 & 3 &&$3 \cdot \alpha + 3/2\cdot L \cdot \beta$& 0.8 s\\
 & \dgxone & 3 & 4 & 4 &&$4 \cdot \alpha + 4/3\cdot L \cdot \beta$& 1.5 s\\
 & \dgxone & 4 & 5 & 5 &&$5 \cdot \alpha + 5/4\cdot L \cdot \beta$& 2.3 s\\
 & \dgxone & 5 & 6 & 6 &&$6 \cdot \alpha + 6/5\cdot L \cdot \beta$& 3.3 s\\
 & \dgxone & 6 & 7 & 7 &Bandwidth&$7 \cdot \alpha + 7/6\cdot L \cdot \beta$& 4.6 s\\
 & \dgxone & 6 & 3 & 7 &Bandwidth&$3 \cdot \alpha + 7/6\cdot L \cdot \beta$& 6.6 s\\
 & \dgxone & 2 & 2 & 3 &Latency&$2 \cdot \alpha + 3/2\cdot L \cdot \beta$& 0.9 s\\
\hline
\allreduce & \dgxone & 8  &4  &4&Latency&$4 \cdot \alpha + 1/2\cdot L \cdot \beta$& 0.3 s\\
  & \dgxone & 16 &6  &6&&$6 \cdot \alpha + 3/8\cdot L \cdot \beta$& 0.6 s\\
  & \dgxone & 24 &8  &8&&$8 \cdot \alpha + 1/3\cdot L \cdot \beta$& 1.3 s\\
  & \dgxone & 32 &10 &10&&$10 \cdot \alpha + 5/16\cdot L \cdot \beta$& 2.9 s\\
  & \dgxone & 40 &12 &12&&$12 \cdot \alpha + 3/10\cdot L \cdot \beta$& 5.6 s\\
  & \dgxone & 48 &14 &14&Bandwidth&$14 \cdot \alpha + 7/24\cdot L \cdot \beta$& 12.8 s\\
  & \dgxone & 48 &6  & 14&Bandwidth&$6 \cdot \alpha + 7/24\cdot L \cdot \beta$& 23.0 s\\
  & \dgxone & 16 &4  &6&Latency&$4 \cdot \alpha + 3/8\cdot L \cdot \beta$& 0.8 s\\
\hline
\broadcast & \dgxone & 2 & 2 & 2 &Latency&$2 \cdot \alpha + 1\cdot L \cdot \beta$& 0.1 s\\
(\reduce) & \dgxone & 6 & 3 & 3 &&$3 \cdot \alpha + 1/2\cdot L \cdot \beta$& 0.3 s\\
 & \dgxone & 12 & 4 & 4 &&$4 \cdot \alpha + 1/3\cdot L \cdot \beta$& 1.0 s\\
 & \dgxone & 18 & 5 & 5 &&$5 \cdot \alpha + 5/18\cdot L \cdot \beta$& 8.5 s\\
 & \dgxone & 6 & 3 & 5 &&$3 \cdot \alpha + 5/6\cdot L \cdot \beta$& 0.9 s\\
\hline
\gathercoll & \dgxone & 1 & 2 & 2 &Latency&$2 \cdot \alpha + 2\cdot L \cdot \beta$& 0.3 s\\
(\scatter) & \dgxone & 2 & 3 & 3 &&$3 \cdot \alpha + 3/2\cdot L \cdot \beta$& 0.9 s\\
  & \dgxone & 3 & 4 & 4 &&$4 \cdot \alpha + 4/3\cdot L \cdot \beta$& 1.6 s\\
  & \dgxone & 4 & 5 & 5 &&$5 \cdot \alpha + 5/4\cdot L \cdot \beta$& 2.7 s\\
  & \dgxone & 5 & 6 & 6 &&$6 \cdot \alpha + 6/5\cdot L \cdot \beta$& 3.8 s\\
  & \dgxone & 6 & 7 & 7 &Bandwidth&$7 \cdot \alpha + 7/6\cdot L \cdot \beta$& 6.0 s\\
  & \dgxone & 6 & 3 & 7 &Bandwidth&$3 \cdot \alpha + 7/6\cdot L \cdot \beta$& 11.4 s\\
  & \dgxone & 2 & 2 & 3 &Latency&$2 \cdot \alpha + 3/2\cdot L \cdot \beta$& 1.0 s\\
\hline
\alltoall & \dgxone & 8 & 3 & 3 &&$3 \cdot \alpha + 3/8\cdot L \cdot \beta$& 2.6 s\\
  & \dgxone & 8 & 2 & 3 &Latency&$2 \cdot \alpha + 3/8\cdot L \cdot \beta$& 3.0 s\\
  & \dgxone & 24 & 8 & 8 &Bandwidth&$8 \cdot \alpha + 1/3\cdot L \cdot \beta$& 133.7 s\\
  & \dgxone & 24 & 2 & 8 &Both&$2 \cdot \alpha + 1/3\cdot L \cdot \beta$& 24.3 s\\
\bottomrule
\end{tabularx}
\caption{\dgxone collectives with chunks ($\chunk$), steps ($\steps$) and rounds ($\rounds$). Time includes both encoding and solving. For \reducescatter and \scatter $\chunk$ should be multiplied by 8.}
\label{fig:dgxone:syn}
\end{table}

\begin{table}
  \begin{tabularx}{\columnwidth}{@{}lHlllXHr@{}}
\toprule
Collective & Topology & $\chunk$ & $\steps$ & $\rounds$ & Optimality & Running Time & Time \\
\midrule
\allgather & \amd & 1 & 4 & 4 &Latency&$4 \cdot \alpha + 4\cdot L \cdot \beta$& 0.5 s\\
(\reducescatter) & \amd & 2 & 7 & 7 &Bandwidth&$7 \cdot \alpha + 7/2\cdot L \cdot \beta$& 1.3 s\\
 & \amd & 2 & 4 & 7 &Both&$4 \cdot \alpha + 7/2\cdot L \cdot \beta$& 1.7 s\\
\hline
\allreduce & \amd & 8 &8&8&Latency&$8 \cdot \alpha + 1\cdot L \cdot \beta$& 0.4 s\\
 & \amd & 16 &14&14&Bandwidth&$14 \cdot \alpha + 7/8\cdot L \cdot \beta$& 0.9 s\\
 & \amd & 16 &8&14&Both&$8 \cdot \alpha + 7/8\cdot L \cdot \beta$& 1.6 s\\
\hline
\broadcast & \amd & 2 & 4 & 4 &Latency&$4 \cdot \alpha + 2\cdot L \cdot \beta$& 0.1 s\\
(\reduce) & \amd & 4 & 5 & 5 &&$5 \cdot \alpha + 5/4\cdot L \cdot \beta$& 0.2 s\\
 & \amd & 6 & 6 & 6 &&$6 \cdot \alpha + 1\cdot L \cdot \beta$& 0.3 s\\
 & \amd & 8 & 7 & 7 &&$7 \cdot \alpha + 7/8\cdot L \cdot \beta$& 0.5 s\\
 & \amd & 10 & 8 & 8 &&$8 \cdot \alpha + 4/5\cdot L \cdot \beta$& 0.6 s\\
\hline
\gathercoll & \amd & 1 & 4 & 4 &Latency&$4 \cdot \alpha + 4\cdot L \cdot \beta$& 0.4 s\\
(\scatter) & \amd & 2 & 4 & 7 &Both&$4 \cdot \alpha + 7/2\cdot L \cdot \beta$& 1.8 s\\
\hline
\alltoall & \amd & 8 & 4 & 8 &Both&$4 \cdot \alpha + 1\cdot L \cdot \beta$& 8.2 s\\
\bottomrule
\end{tabularx}
\caption{\amd collectives with chunks ($\chunk$), steps ($\steps$) and rounds ($\rounds$). Time includes both encoding and solving. For \reducescatter and \scatter $\chunk$ should be multiplied by 8.}
\label{fig:amd:syn}
\end{table}

\subsubsection{Synthesizing All Collectives}
Collective communication libraries need to support a large and diverse
set of hardware architectures.  Efficiently implementing latency and
bandwidth optimal algorithms for various topologies is
time-consuming and error-prone. \tool's synthesis based approach
allows it to easily extend the set of algorithms through search:
\tool{} synthesizes algorithms for \alltoall, \gathercoll and \scatter
where no such counterparts exist in NCCL.

\subsubsection{Synthesis time}
The longest synthesis time is just over 2 minutes and most of the time
under 10 seconds.  The synthesis problem is non-trivial and its complexity is
defined by both the collective, as well as the hardware topology we synthesize
for. The clever encoding described in Section~\ref{sec:encoding} was critical
for achieving these fast synthesis times. As a point of comparison, synthesizing
the 24-chunk 8-step bandwidth-optimal \alltoall algorithm with a more direct
encoding with a Boolean variable for each tuple $(c, n, n', s) \in \sends$ did
not finish within 60 minutes. With the better encoding the synthesis finishes in
just over 2 minutes.  

\subsection{Performance Evaluation}
In this section, we compare \tool's generated algorithms with NCCL and
RCCL on the NVIDIA and AMD hardware.  Our code generation uses a protocol similar to 
the simple protocol (i.e., NCCL\_PROTO=Simple). Thus, we use NCCL with the simple protocol as our baseline. 
We investigate the performance of \allgather, \allreduce, and \alltoall
as they are popular primitives in different workloads including machine learning. For each
hardware platform and collective, we generate multiple algorithms; for
each algorithm, we lower using (1) a single kernel-launch, or (2) 
multiple \texttt{cudaMemcpy} calls with one per step. Each algorithm uses a push model for copying and
when \tool{} is compared with NCCL, we exhaustively search the size and the number of thread blocks and
report the best performing combination for both \tool{} and NCCL. 
See Section~\ref{sec:lowering} for more details.

\begin{figure*}[h]
  \centering
  \begin{subfigure}[h]{0.5\textwidth}
    \includegraphics[page=1,width=\columnwidth]{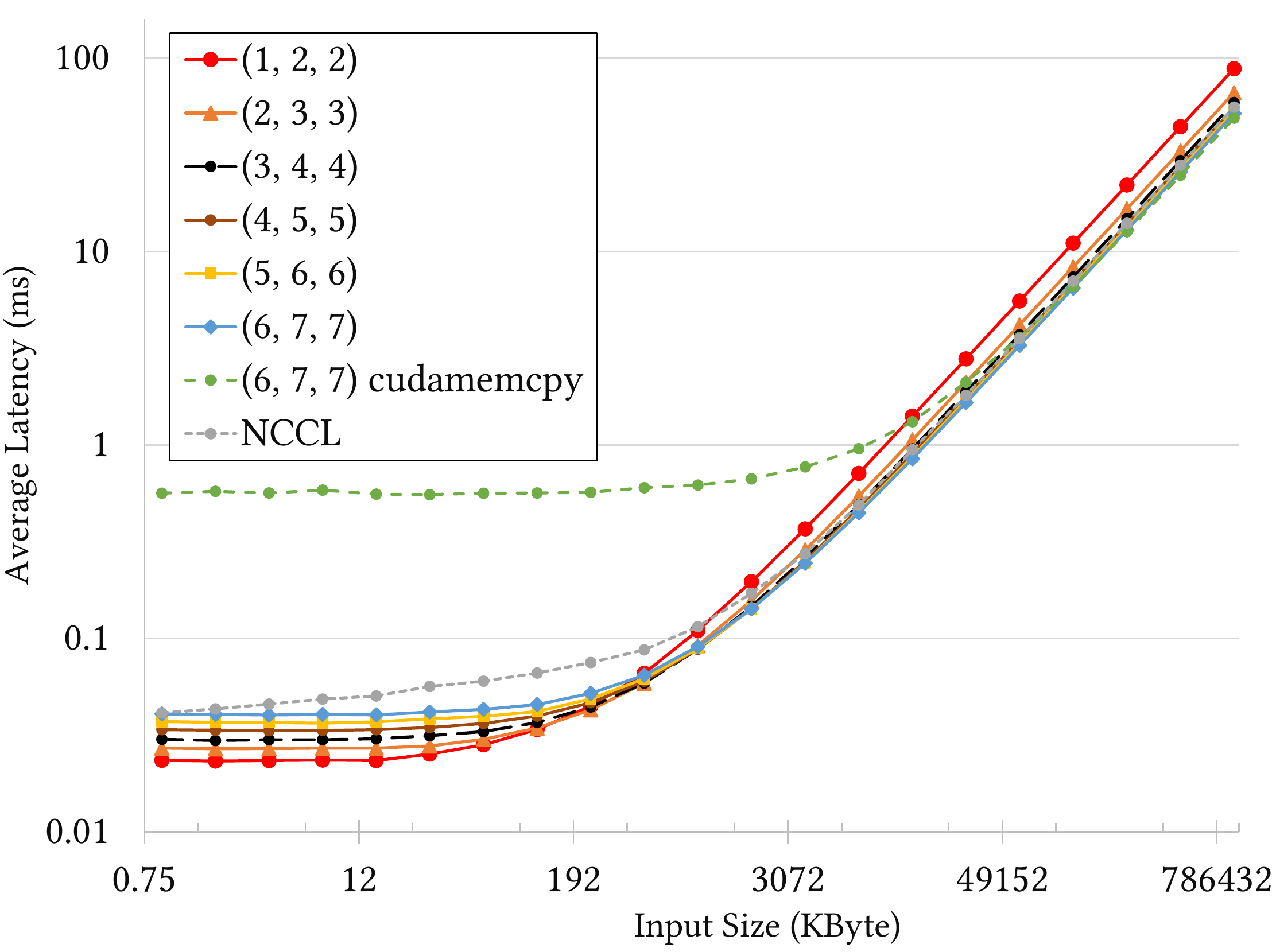}
    \caption{Average latency}
    \label{fig:dgx1-res-allgather-avglat}
  \end{subfigure}%
  \begin{subfigure}[h]{0.5\textwidth}
    \includegraphics[page=2,width=\columnwidth]{figures/evals-camera-ready}
    \caption{Speedup}
    \label{fig:dgx1-res-allgather-speedup}
  \end{subfigure}%
  \caption{\allgather performance comparison with NCCL}
  \label{fig:dgx1-res-allgather}
\end{figure*}
Figure~\ref{fig:dgx1-res-allgather} compares \tool's generated code for \allgather with NCCL's \allgather.
A point on Figure~\ref{fig:dgx1-res-allgather-avglat} ($x$,$y$) shows the
running time in $y$ milliseconds as a function of send input
buffer size in $x$ Kbytes while a point on Figure~\ref{fig:dgx1-res-allgather-speedup}
shows the $y$ speedup of \tool's generated code over NCCL's \allgather as a function of send input
buffer size in $x$ Kbytes. We plot one line per algorithm denoted as
$(\chunk, \steps, \rounds)$ for respectively chunks, steps, and rounds as defined in Table~\ref{fig:dgxone:syn}.
To show the impact of our lowering, we plot two versions of a bandwidth
optimal algorithm $(6,7,7)$ (which utilizes a push-copy) and $(6,7,7)$
\texttt{cudaMemcpy}.  The latter of which shows the significant impact lowering
can have on the performance. To simplify the figure, we only show algorithms that were faster on at least one input size we experimented with. 
As it can be seen from the lines, \tool{} is up-to $2.2\times$ faster than NCCL's \allgather on small sizes and $1.14\times$ faster on larger sizes.
It is possible for \tool{} to automatically switch between multiple implementations based on the input size. In which case, \tool{} will consistently outperform NCCL.  

\begin{figure*}[h]
  \centering
  \begin{subfigure}[h]{0.5\textwidth}
    \includegraphics[page=3,width=\columnwidth]{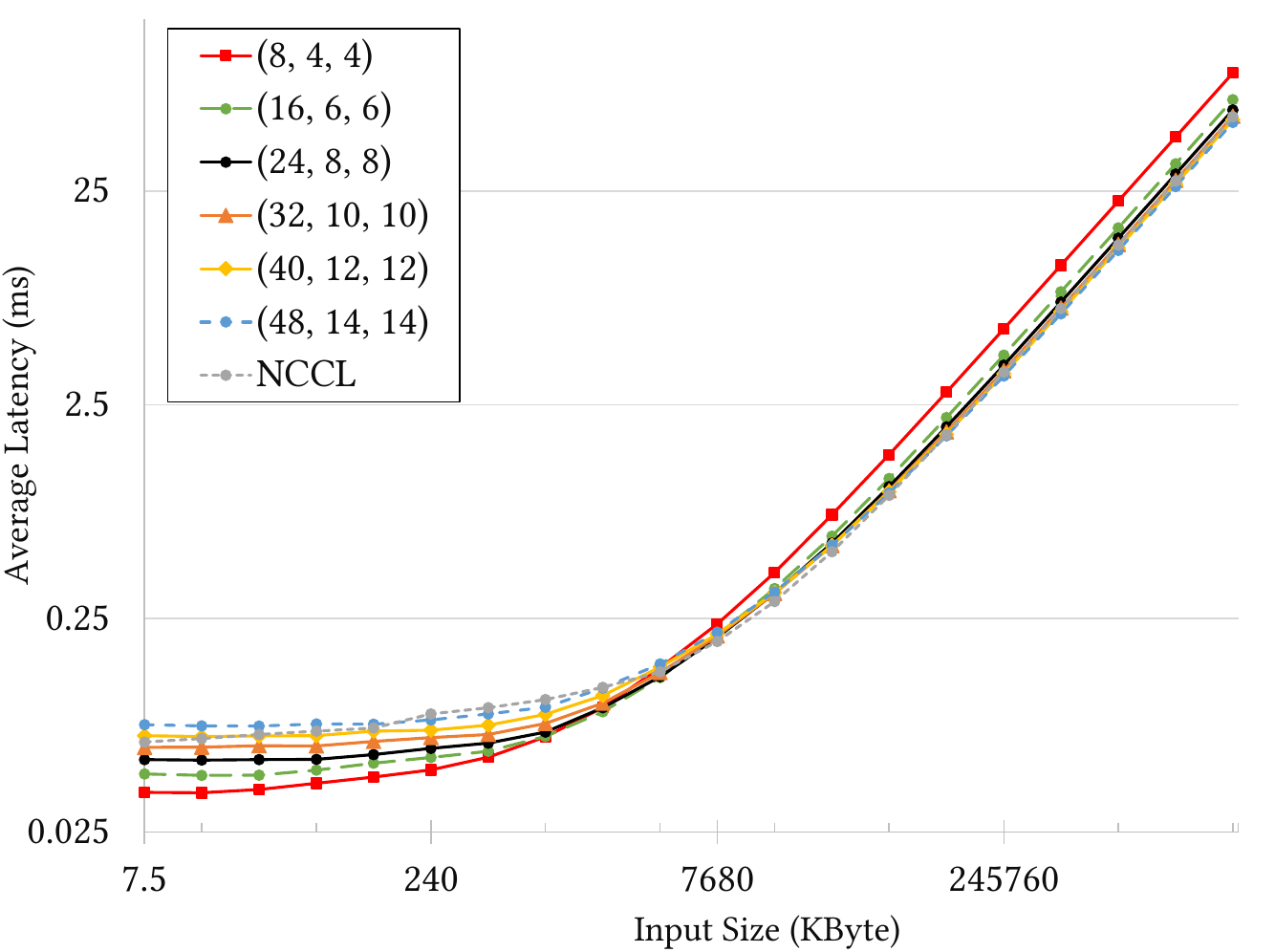}
    \caption{Average latency}
    \label{fig:dgx1-res-allreduce-avglat}
  \end{subfigure}%
  \begin{subfigure}[h]{0.5\textwidth}
    \includegraphics[page=4,width=\columnwidth]{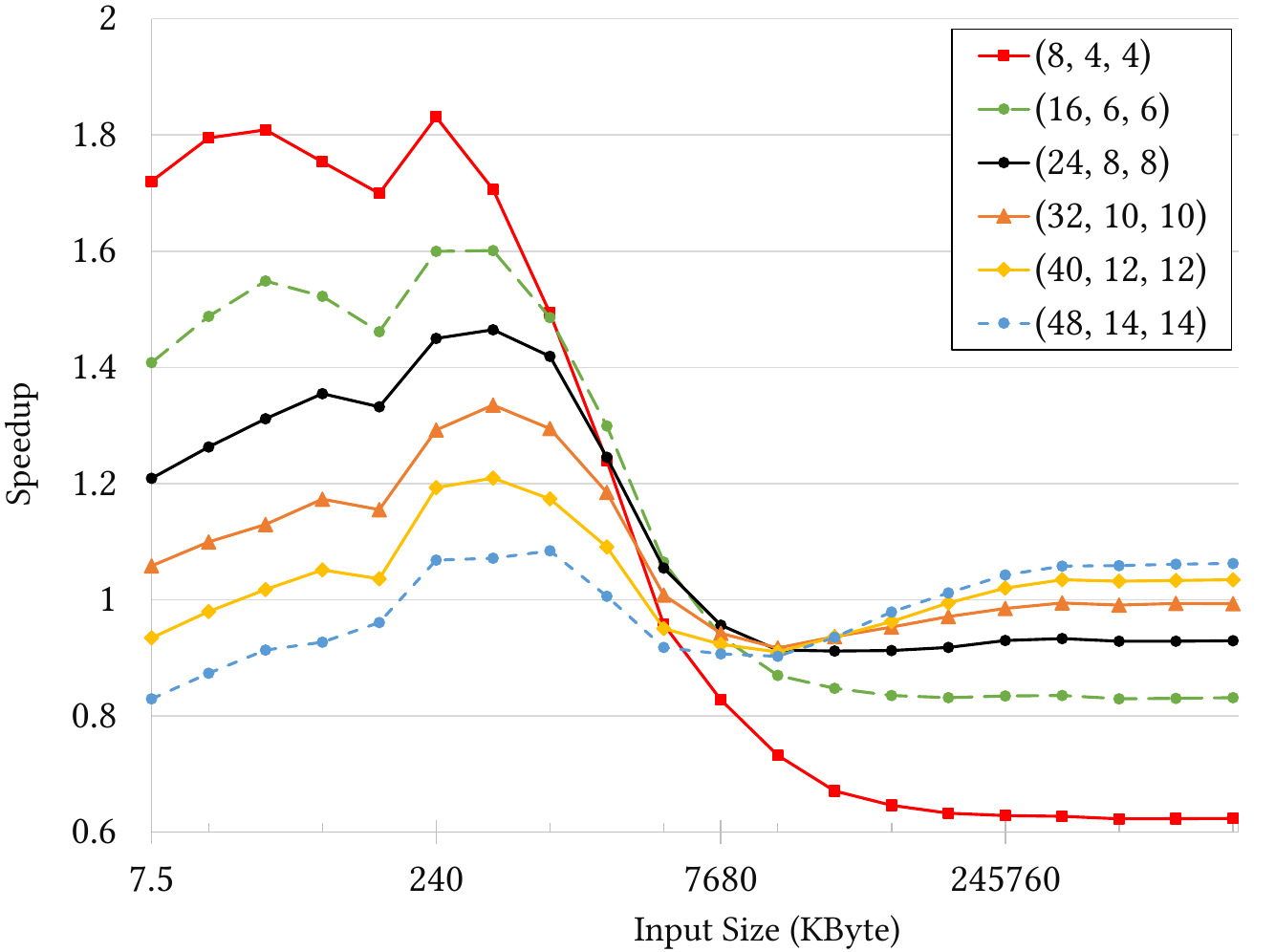}
    \caption{Speedup}
    \label{fig:dgx1-res-allreduce-speedup}
  \end{subfigure}%
  \caption{\allreduce performance comparison with NCCL}
  \label{fig:dgx1-res-allreduce}
\end{figure*}

\begin{figure*}[h]
  \centering
  \begin{subfigure}[h]{0.5\textwidth}
    \includegraphics[page=5,width=\columnwidth]{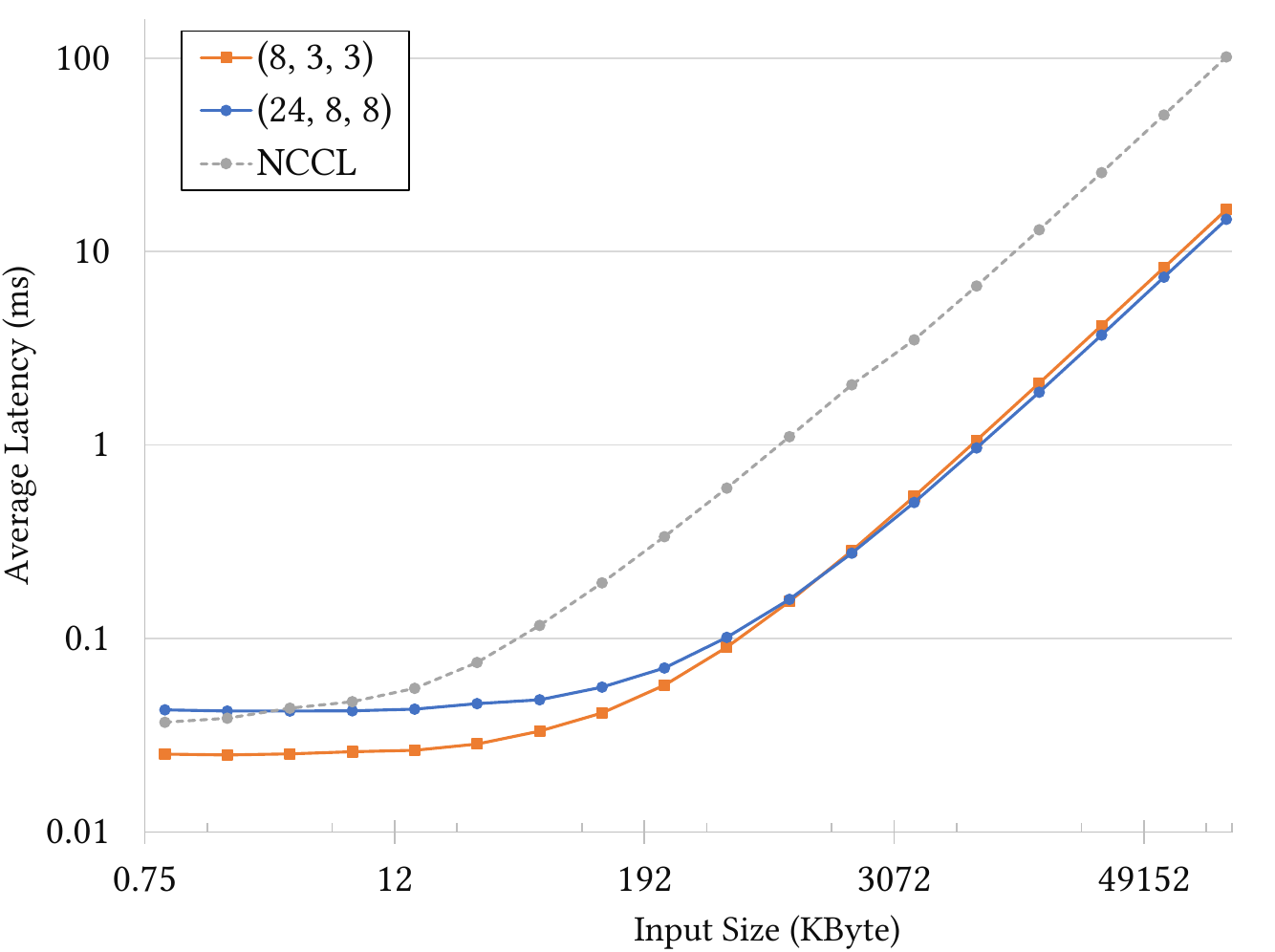}
    \caption{Average latency}
    \label{fig:dgx1-res-alltoall-avglat}
  \end{subfigure}%
  \begin{subfigure}[h]{0.5\textwidth}
    \includegraphics[page=6,width=\columnwidth]{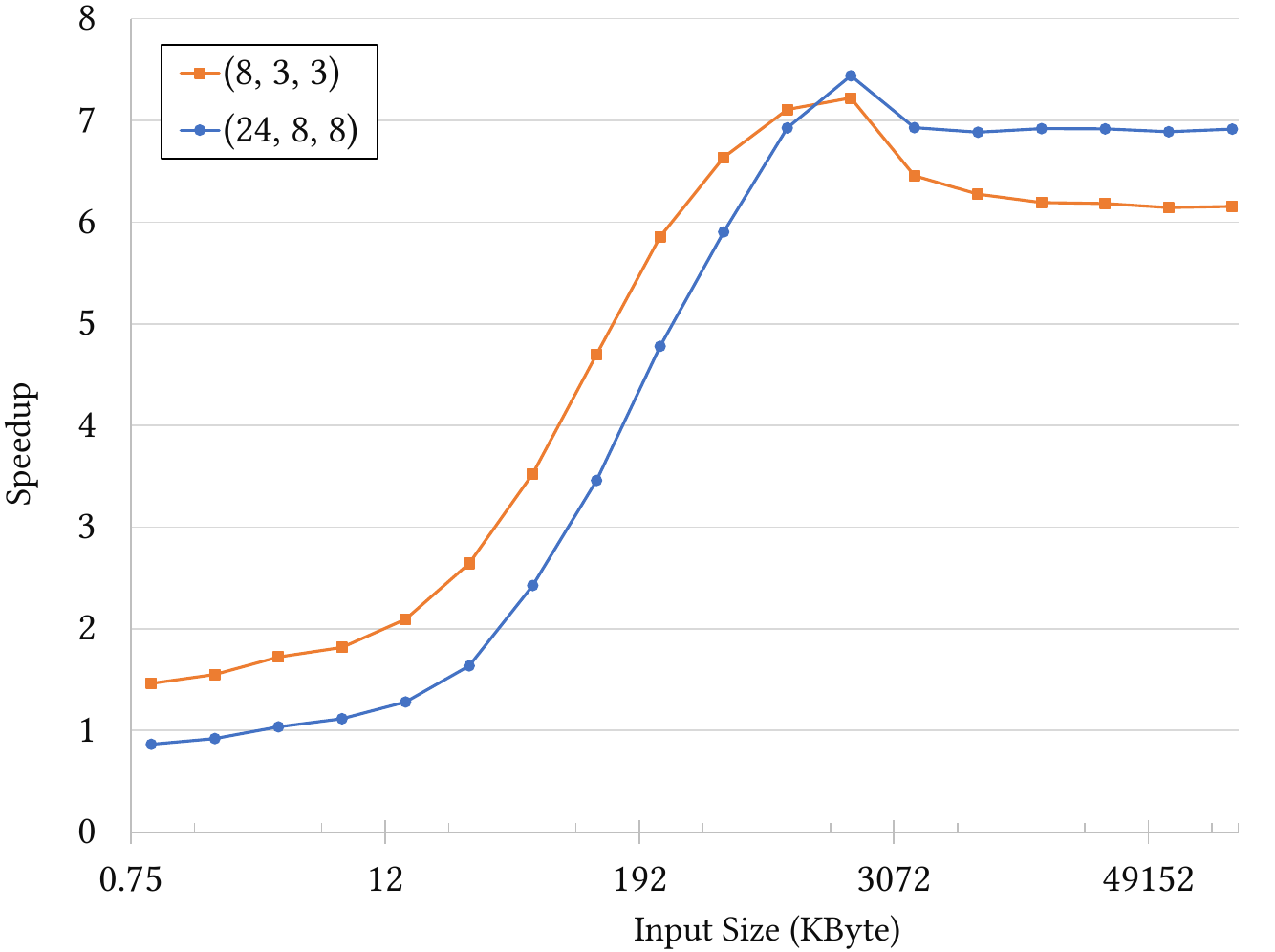}
    \caption{Speedup}
    \label{fig:dgx1-res-alltoall-speedup}
  \end{subfigure}%
  \caption{\alltoall performance comparison with NCCL}
  \label{fig:dgx1-res-alltoall}
\end{figure*}

Likewise, Figure~\ref{fig:dgx1-res-allreduce} shows the running time in milliseconds (Figure~\ref{fig:dgx1-res-allreduce-avglat}) or speedup (Figure~\ref{fig:dgx1-res-allreduce-speedup}) 
for \allreduce as a function of the receive input size.  Each line denotes
$(\chunk, \steps, \rounds)$ for respectively chunks, steps, and rounds, respectively. With the exception of $4$ middle sizes,
\tool{} beats NCCL's \allreduce with an 8-chunk algorithm for small input sizes by up-to $1.8\times$ and with a 48-chunk algorithm for large input sizes by up-to $1.06\times$.

\alltoall is a complex algorithm which is very difficult to write
efficiently by hand. Unlike the prior collectives, NCCL does not
natively support \alltoall; instead, NCCL suggests using $N$
point-to-point exchanges (for $N$ GPUs) and thus its resulting
algorithm is neither bandwidth nor latency optimal.  Because \tool{}
uses program synthesis to generate optimal algorithms, it is able to
synthesize three \alltoall{} algorithms in a matter of minutes.
Figure~\ref{fig:dgx1-res-alltoall-avglat} shows the latency in
milliseconds of \tool{} and NCCL as a function of input size while
Figure~\ref{fig:dgx1-res-alltoall-speedup} shows speedup over NCCL
again as a function of input size.  Each line denotes
$(\chunk, \steps, \rounds)$ for respectively chunks, steps, and rounds
and demonstrates a speedup of over $6.8\times$ for large input sizes and
over $1.4\times$ for small input sizes, depending on whether we pick a
latency or bandwidth optimal implementation from \tool.  This
significant speedup really shows off the power of \tool's automated
approach to building algorithms tailored specifically to a hardware
architecture.

\begin{figure*}[h]
  \centering
  \begin{subfigure}[h]{0.5\textwidth}
    \includegraphics[page=7,width=\columnwidth]{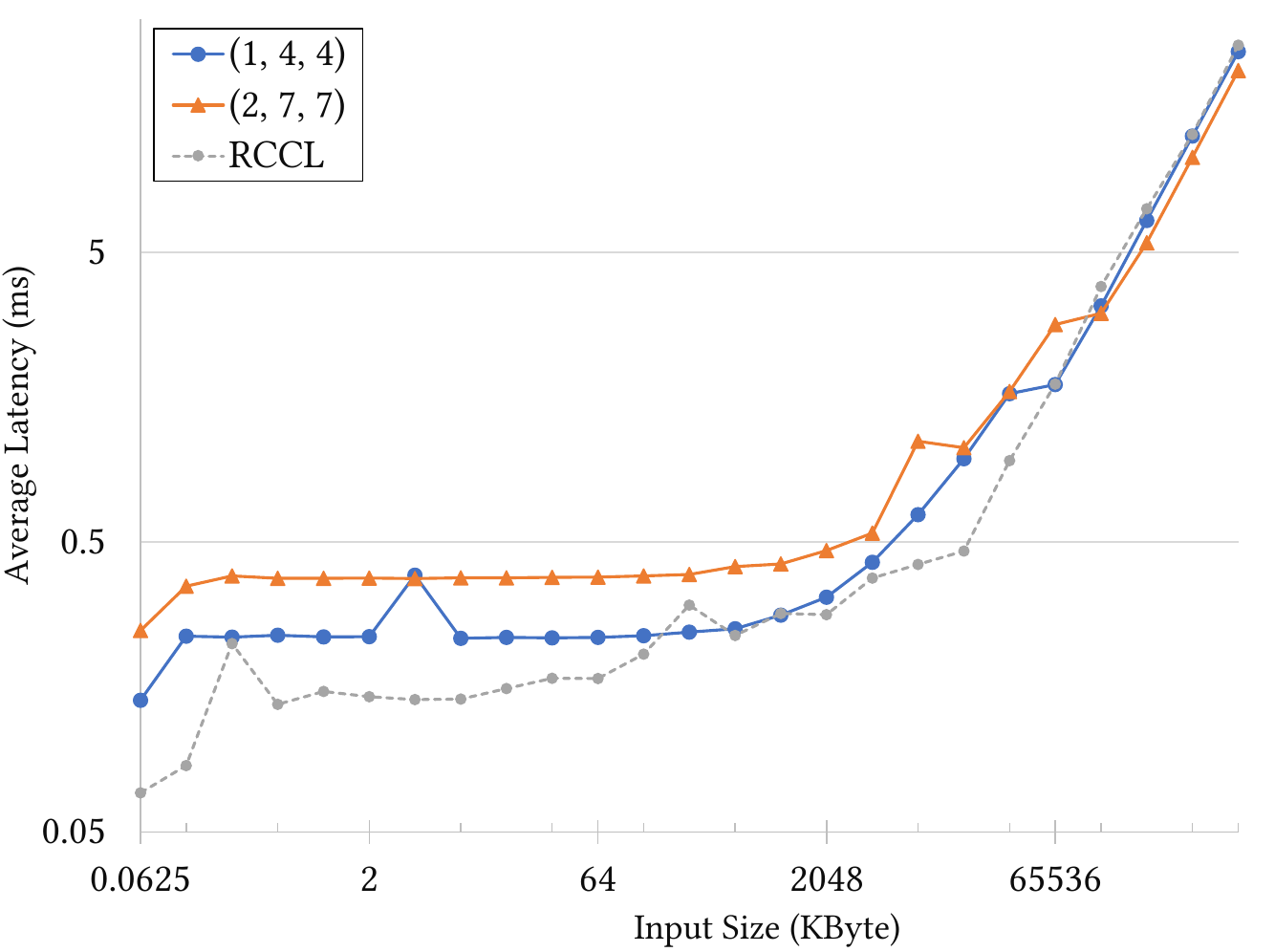}
    \caption{Average latency}
    \label{fig:amd-res-allgather-avglat}
  \end{subfigure}%
  \begin{subfigure}[h]{0.5\textwidth}
    \includegraphics[page=8,width=\columnwidth]{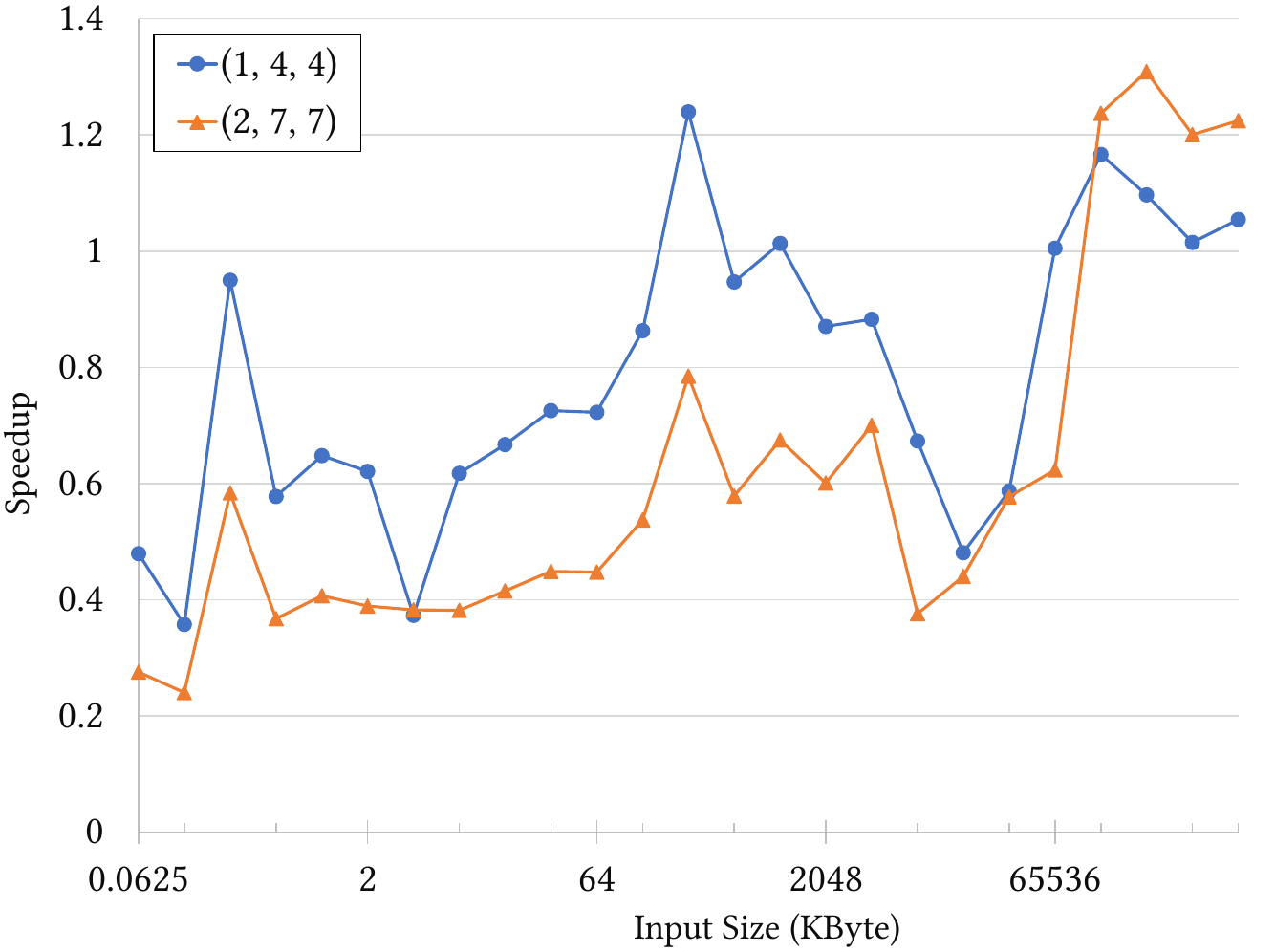}
    \caption{Speedup}
    \label{fig:amd-res-allgather-speedup}
  \end{subfigure}%
  \caption{\allgather performance comparison with RCCL}
  \label{fig:amd-res-allgather}
\end{figure*}
Lastly, we demonstrate \allgather on the Gigabyte AMD workstation.
Like the other plots, a point on Figure~\ref{fig:amd-res-allgather}
($x$,$y$) shows the latency or speedup $y$ for \allgather as a function
of the receive input size in bytes $x$.  We plot two algorithms,
$(1,4,4)$ and $(2,7,7)$; it is clear that (i) the lower latency
algorithm $(1,4,4)$ is better at smaller input sizes, (ii) the higher
bandwidth algorithm $(2,7,7)$ is faster for large input sizes, and
(iii) \tool's generated code is faster than RCCL for large sizes but
slower for medium and small sizes. The Gigabyte machine, in
particular, is new hardware and \tool{} can synthesize new algorithms
and implementations for it; this shows \tool{} can help design future
interconnects and co-design them with communication libraries.

These graphs in concert show that \tool{} is able to synthesize
algorithms along the Pareto-optimal frontier and also lower than to
hardware so as to be competitive with a hand optimized baseline.

\section{Related Work}
The message passing interface (MPI)~\cite{dongarra2013mpi} is a widely-used standardized abstraction for communication primitives in a multi processor system. Implementations of MPI provide reliable and portable implementations of collective primitives. Efficient algorithms for implementing these primitives is a long-studied research area~\cite{pjevsivac2007performance, chan2007collective, thakur2005optimization}, including optimized algorithms for specific architectures like mesh, hypercube, or fat-tree\cite{scott1991efficient,bokhari1992complete,barnett1993global} and for clusters of shared-memory processors~\cite{sistare1999optimization,traff2002improved,sanders2002hierarchical,tipparaju2003fast}. The class of $k$-synchronous algorithms studied in this paper is designed to include many of the algorithms proposed in these works and implemented in popular MPI implementations such as MPICH~\cite{thakur2005optimization} and OpenMPI~\cite{gabriel2004open}.

We evaluated OpenMPI, either through builtin CUDA capability or through Unified Communication X~(UCX)~\cite{ucx}.
They lack custom implementations for architectures such as the \dgxone{}, and result in subpar performance compared with our NCCL baselines.
NCCL~\cite{nccl} is a library for multi NVIDIA GPU systems and it utilizes the underlying hardware transport such as NVLink, NVSwitch or Infiniband for an efficient implementation of collective primitives. RCCL~\cite{rccl} is a port of NCCL for AMD GPUs and the HIP compiler suite. While these libraries provide efficient implementations for a limited set of algorithms, \tool{} is able to synthesize a wide range of algorithms suitable for different input sizes and generate collective primitives that are not even a part of standard MPI set.

There are also hybrid algorithms~\cite{barnett1994building, chan2007collective} that switch between latency- and bandwidth-optimal algorithm along each dimension of a mesh network. However, to the best of our knowledge, these prior works do not seek to identify algorithms that are Pareto-optimal for a given topology. In contrast to these prior works, the goal of this paper is to automatically synthesize Pareto-optimal algorithms for a given topology.  

There are also hierarchical approaches to implement collective primitives in distributed systems. Horovod~\cite{alex2018horovod} implements collective primitives by using NCCL locally in node and MPI across nodes. Others such as BlueConnect~\cite{blueconnect} and PLink~\cite{plink} exploit the hierarchical network topology of a cloud system or a data center to improve the performance of collective primitives. In this paper, we focus on synthesizing algorithms for a single node with multiple GPU, while the above approaches are beneficial on multi node systems.

Motivated by recent resurgence in machine-learning workloads, recent research has focused on optimizing the communication of distributed machine learning. Blink~\cite{wang2020blink}, the closest to our work, automatically synthesizes bandwidth-efficient collective primitives for a given topology. This work is based on packing spanning trees and is suitable for one-to-many collective primitives such as broadcast and reduce, and implements \allreduce as a reduce followed by a broadcast. Blink is not guaranteed to generate bandwidth-optimal algorithms as it heuristically selects a few trees based on an approximate spanning-tree packing algorithm. Moreover, Blink's focus is not on generating latency-optimal algorithms. In contrast, this work generates latency- and bandwidth-optimal algorithms for a given topology. There are also other works~\cite{zhang2017poseidon,hashemi2019tictac,jayarajan2019priority,peng2019generic} on optimizing distributed machine learning that do so by overlapping computation and communication and are orthogonal to this work. 




\section{Conclusion}
This paper introduces \tool: a systematic method to synthesize
algorithms in the Pareto-frontier spanning from the latency-optimal
algorithm to the bandwidth-optimal algorithm for a given collective on
an input topology. We characterize a class of algorithms that captures
a broad set of known algorithms and prove Pareto-optimality of both
known algorithms and synthesized new algorithms. We automatically
generate an implementation of these algorithms that is competitive
with manually hand-tuned communication kernels in use today.

\bibliographystyle{ACM-Reference-Format}
\bibliography{references}


\end{document}